\documentclass[12pt]{article}
\usepackage[top=1in, bottom=1in, left=1.25in, right=1.25in]{geometry}
\usepackage{float}
\usepackage{color}
\usepackage{tikz}
\usepackage{setspace}
\usepackage[toc,page]{appendix}
\usepackage{amssymb}
\usepackage{showlabels}
\usepackage[normalem]{ulem}
\usepackage{multirow}
\usepackage{tikz}
\usepackage[toc]{appendix}
\usepackage{chngcntr}
\usetikzlibrary{shapes,arrows}
\usepackage{subfigure}
\usepackage[round]{natbib}
\usepackage{latexsym}
\usepackage{amsmath}
\usepackage{graphicx}
\usepackage{caption}
\providecommand{\keywords}[1]{\textbf{\textit{Key words:}} #1}
\usepackage{bm}
\usepackage[pdftex,hypertexnames=false,linktocpage=true]{hyperref}
\hypersetup{colorlinks=true,linkcolor=blue,anchorcolor=blue,citecolor=blue,filecolor=blue,urlcolor=blue,bookmarksnumbered=true,pdfview=FitB}

\newcommand{\biblist}{\begin{list}{}
		{\listparindent 0.0cm \leftmargin 0.50cm \itemindent -0.50 cm
			\labelwidth 0 cm \labelsep 0.50 cm
			\usecounter{list}}\clubpanelty4000\widowpanelty4000}
	\newcommand{\ebiblist}{\end{list}}

\def\T{{ \mathrm{\scriptscriptstyle T} }}

\newtheorem{theorem}{Theorem}

\newtheorem{remark}{Remark}

\def\boxit#1{\vbox{\hrule\hbox{\vrule\kern6pt\vbox{\kern6pt#1\kern6pt}\kern6pt\vrule}\hrule}}

\title{\LARGE\bf Bayesian Sparse Propensity Score Estimation for Unit Nonresponse }
\author{Hejian Sang \and  Gyuhyeong Goh \and Jae Kwang Kim}

\begin{document}
\baselineskip .3in
\maketitle

\begin{abstract}
Nonresponse weighting adjustment using propensity score is a popular method for handling unit nonresponse. However, including all available auxiliary variables into the propensity model can lead to inefficient and inconsistent estimation, especially with high-dimensional covariates.  In this paper, a new Bayesian method using the Spike-and-Slab prior  is proposed for sparse propensity score estimation. The proposed method is not based on any model assumption on the outcome variable and is computationally efficient. Instead of doing model selection and parameter estimation separately as in many frequentist methods, the proposed method simultaneously selects the sparse response probability model and provides consistent parameter estimation. Some asymptotic properties of the proposed method are presented. The efficiency of this sparse propensity score estimator is further improved by incorporating related auxiliary variables from the full sample. The finite-sample performance of the proposed method is investigated in two limited simulation studies, including a partially simulated real data example from the Korean Labor and Income Panel Survey.
\end{abstract}


\newpage
\section{Introduction}\label{sec::Intro}

Nonresponse in the collected data is a common problem in survey sampling, clinical trials, and many other areas of research. Ignoring nonresponse can lead to biased estimation unless the  response mechanism is completely missing at random \citep{rubin1976inference}. To handle nonresponse, various statistical methods have been developed. \cite{little2014statistical} and \cite{kim2013statistical} provide  comprehensive overviews of the statistical methods for handling missing data.

The propensity score weighting is one of the most popular tools for adjusting for nonresponse bias, which builds on a model for the response probability only and uses the inverse of the estimated response probabilities as weights for estimating parameters. The propensity score weighting method is well established in the literature. See \cite{rosenbaum1987model}, \cite{flanders1991analytic}, \cite{robins1994estimation}, \cite{robins1995analysis}, \cite{paik1997generalized} and \cite{kim2007nonresponse}.  However, when the dimension of the covariates for the propensity score is high, the full response model including all the covariates may have several problems. First, the computation for parameter estimation can be problematic as it involves high dimensional matrix inversion and the convergence is not guaranteed. Second, estimating zero coefficients in the propensity model increases the variability of the propensity scores and thus leads to inefficient estimation of the model parameters. Furthermore, the asymptotic normality of the propensity score estimator is not guaranteed if the dimension of the covariates is high.  That is, the assumptions for the Central Limit Theorem may not be satisfied if we include all the covariates into the propensity model. Therefore, model selection to obtain a sparse propensity model is a challenging but important practical problem.  While sparse model estimation is well studied in the literature \citep{Tibshirani:1996,Fan:2001, Zou:2005,Zou:2006,Park:2008,Kyung:2010}, to the best of our knowledge, not much work has been done for sparse propensity score estimation in the missing data context.


Our main goal is to develop a valid inference procedure for estimating parameters with the sparse propensity score adjustment in a high dimensional setup. In this paper, we propose a new Bayesian approach for sparse propensity score estimation. One advantage of the Bayesian approach is that both model selection and parameter estimation can be simultaneously performed in the posterior inference. To develop the sparse posterior distribution, we use stochastic search variable selection with the Spike-and-Slab prior, which is a mixture of flat distribution and degenerate distribution at zero, or a mixture of their approximations \citep{mitchell1988bayesian, george1993variable, george1997approaches,narisetty2014bayesian}. However, implementing the Bayesian variable selection method to  propensity score (PS) estimation is challenging, because the likelihood function for the parameter of interest is not available as the outcome model is unspecified. To resolve this issue, we derive an approximate likelihood from the sampling distribution of the PS estimator before applying Spike-and-Slab prior for the PS model selection.  Note that, selecting a correct propensity model does not necessarily achieve efficient estimation.
Incorporating auxiliary variables observed from the full sample \citep{zhou2012efficient} using generalized method of moments technique, however, can achieve optimal estimation.
Thus, to achieve the optimal PS estimation in a Bayesian way, we propose using a working outcome model and the Spike-and-Slab prior to select only relevant auxiliary  variables.
The proposed Bayesian method is implemented by data augmentation algorithm \citep{tanner1987calculation, wei1990monte} and the computation of posterior distribution is fast and efficient.

The rest of this paper is organized as follows. In Section \ref{sec::setup}, we introduce the basic setup of the PS estimation. The proposed method is fully described in Section \ref{sec:proposed}. Some asymptotic theories including model selection consistency are established in Section \ref{sec::asymp}. The optimal sparse PS estimator is introduced in Section \ref{sec::extension}. The performance of the proposed method is examined through extensive simulation studies in Section \ref{sec::simulation}. Some concluding remarks are made in Section \ref{sec::discussion}. All technical proofs are relegated to Appendix.

\section{Setup}\label{sec::setup}


Let $(x_1, y_1),(x_2, y_2),
\ldots,(x_n, y_n)$ be $n$ independent and identically distributed (IID) realizations  of random vector $(X,Y)$, where $Y$ is a scalar response variable and $X$ is a $p$-dimensional vector of covariates. The dimension $p$ is allowed to increase with the sample size. Suppose we are interested in estimating parameter $\theta=E(Y)$, which  can be estimated by $\hat \theta_n=n^{-1}\sum_{i=1}^n y_i$, under complete response. Note that no distribution assumptions are made on $(X,Y)$.

To handle the missing data problem, the response propensity model can be used. To introduce this PS method, suppose that $x_i$ are fully observed and $y_i$ are subject to missingness. Let $\delta_i$ be the response indicator of $y_i$, that is,
\begin{eqnarray}
\delta_i=\left\lbrace \begin{array}{ll}
1 & \text{if $y_i$ is observed}\\
0 & \text{if $y_i$ is missing}.
\end{array}\right.\nonumber
\end{eqnarray}
Assume that $\delta_i$ are independently distributed from a Bernoulli distribution with the success probability $\Pr(\delta_i=1|x_i, y_i)$. We further assume that the response mechanism
is missing at random (MAR) in the sense that
\begin{eqnarray}\label{sec2:eq2}
\Pr(\delta=1|X, Y)=\Pr(\delta=1|X).\nonumber
\end{eqnarray}
Furthermore, we assume a parametric model for the response probability
\begin{eqnarray}\label{sec2:eq3}
\Pr(\delta_i|X)=\pi(\phi;X)=G\left(X^\T\phi\right),\label{phi}
\end{eqnarray}
where $G:\mathbb{R} \to [0,1]$ is a known distribution function and $\phi=(\phi_1,\phi_2,\ldots,\phi_p)^\T$ is a $p$-dimensional unknown parameter. Then the propensity score estimator of $\theta$, say $\hat{\theta}_{\text{PS}}$, can be obtained by solving
\begin{eqnarray}\label{sec2:eq4}
U_{\text{PS}}( \theta,\hat \phi)=\sum_{i=1}^{n}\frac{\delta_i}{\pi(\hat\phi;x_i)} (y_i-\theta)=0,
\end{eqnarray}
with respect to $\theta$, where $\hat{\phi}$ is a consistent estimator of $\phi$ in (\ref{phi}). From the response model in (\ref{sec2:eq3}), the maximum likelihood estimator (MLE) of $\phi$ is obtained by
maximizing the log-likelihood function,
\begin{eqnarray}\label{sec2:eq5}
l_n(\phi)=\sum_{i=1}^n\log f(\delta_i|x_i;\phi),\label{eq:likelihood_phi}
\end{eqnarray} where  $	f(\delta_i|x_i;\phi)= \left\lbrace \pi(x_i;\phi)\right\rbrace ^{\delta_i}\left\lbrace 1-\pi( x_i;\phi)\right\rbrace ^{1-\delta_i}$ and the score equation for obtaining $\hat{\phi}$ can be written as
\begin{equation}
\label{sec2:eq6}
S_n( \phi) \equiv \frac{ \partial}{\partial \phi} l_n (\phi) = 0.
\end{equation}

However, when $\phi$ is sparse, that is, $\phi$ contains many zero values, the MLE from the fully saturated model  often increases its variance and fails to be consistent \citep{Zou:2006}. Such phenomenon unfavorably leads to poor inference on the parameter of interest $\theta$. In addition, the propensity model with unnecessary covariates may increase the variance of the resulting PS estimator.  However, including important covariates into the PS model is still critical to obtain consistency.

Penalized likelihood estimation techniques have been proposed to overcome the drawbacks of MLE for high dimensional regression  problems. Thus, we may achieve sparse and consistent estimation for $\phi$ by adding a suitable penalty function to (\ref{sec2:eq5}). For example, LASSO \citep{Tibshirani:1996} produces a sparse estimator of $\phi$ via $L_1$-penalization,
\begin{eqnarray}
\hat{\phi}_{\text{LASSO}}=\arg\min_{\phi} \left\{-l_n(\phi)+\lambda \sum_{j=1}^p |\phi_i|\right\},\label{lasso}
\end{eqnarray}
where $\lambda\geq 0$ is a predetermined parameter to control the degree of penalization. Thus, we can easily obtain a penalized PS estimate of $\theta$ by solving (\ref{sec2:eq4}) for a given $\hat{\phi}_{\text{LASSO}}$. However, the penalized likelihood method is limited to the point estimation in the PS method. The derivation of the variance estimator of $\hat{\theta}_{\text{PS}}$ is very challenging under the penalization approach \citep{lee2016exact,tibshirani2016exact}. More importantly, the resulting PS estimator can be inefficient as it does not fully incorporate all available information.  That is, the penalized likelihood estimation technique can only select the covariates in the true PS model, which, as shown in Section \ref{sec::asymp}, does not necessarily lead to efficiency gain. To get efficient PS estimation, it is better to include covariates correlated with $Y$, even if they are not selected in the true PS model.

 All the aforementioned concerns have motivated us to tackle the sparse propensity estimation problem under a Bayesian framework. We use Bayesian stochastic variable search and approximate Bayesian computation \citep{beaumont2002approximate, soubeyrand2015weak} for the sparse propensity score estimation and optimal PS estimation. The details are discussed in the following section.

\section{ Bayesian Sparse Propensity Score Estimation}\label{sec:proposed}

To formulate our proposal, we first introduce the Bayesian PS estimation discussed in \cite{BayesianPS}. Note that $(\hat{\theta}_{PS}, \hat{\phi})$ is the solution to the joint estimating equations in (\ref{sec2:eq4}) and (\ref{sec2:eq6}).
Using  asymptotic normality
\begin{eqnarray}\label{claim1}
n^{-1/2} \left.\left\lbrace \begin{matrix}S_n \left( \phi\right)\\
U_{PS}\left(\theta, \phi  \right)
\end{matrix}\right\rbrace   \right|\left( \phi,\theta \right)  \stackrel{\mathcal{L}}{ \longrightarrow}  N\left(  0, \Sigma \right), \label{eq:asym}
\end{eqnarray}
where
$\Sigma=:\Sigma\left(\phi,\theta \right) $ is a non-stochastic, symmetric, and positive-definite matrix,
as $n\xrightarrow{}\infty$, the approximate posterior distribution of $(\phi, \theta)$ proposed by \cite{BayesianPS} is given by
\begin{eqnarray}
p(\phi, \theta\mid  \mbox{data})\propto g\left\lbrace S_n(\phi), U_{PS}(\theta,\phi)\mid \theta, \phi\right\rbrace p(\theta)p(\phi),
\label{post1}
\end{eqnarray}
where $g\left\lbrace S_n(\phi), U_{PS}(\theta,\phi)\mid \theta, \phi\right\rbrace$ is the density function of the sampling distribution in (\ref{claim1}), $p(\phi)$ and $p(\theta)$ are the prior distributions of $\phi$ and $\theta$, respectively. Note that, in (\ref{post1}), instead of using the explicit likelihood function of $\phi$ in (\ref{eq:likelihood_phi}), the asymptotic distribution of the estimating equations is used to replace the likelihood function as we do not make any distribution assumption on $Y$.

To extend the method of \cite{BayesianPS} to the sparse propensity score model, we introduce a vector of latent variables $z=\left(z_1, z_2, \cdots, z_p \right)^\T $, such that
\begin{eqnarray}
z_j=\left\{
\begin{array}{ll}
1 & \text{if $\phi_j\neq 0$}\\
0 & \text{if $\phi_j=0$}
\end{array}\right.,~j=1,2,\ldots,p.
\end{eqnarray}
Thus, $z_j$ is an indicator function for including the $j$-th covariate into the response probability model in (\ref{phi}).

To account for the sparsity of the response model, we assign the Spike-and-Slab Gaussian mixture prior for $\phi$ and an independent Bernoulli prior for $z$ as follows:
\begin{eqnarray}
\label{ss:prior}\phi_j|z_j&\overset{ind}{\sim}& \text{N}(0,\nu_{0}(1-z_j)+\nu_{1}z_j),\\
\label{ber:prior}z_j&\overset{ind}{\sim} &\text{Ber}(w_j),
\end{eqnarray}
where $w (\in (0,1))$, $\nu_0(>0)$, and $\nu_1(>\nu_0) $ are deterministic hyperparameters. To induce sparsity for $\phi$, the scale hyperparameters $\nu_0$ and $\nu_1$ need to be small and large fixed values, respectively.  For example, we use $\nu_0=10^{-4}$ and $\nu_1=10^4$ in the simulation study in Section \ref{sec::simulation}. The mixing probability $w_j$ can be interpreted as the prior probability that $\phi_j$ is nonzero.  Under the absence of prior information for $\phi$, we can set $w_j=0.5$ for all $j$ or set a uniform prior for $w_j$. Note that (\ref{ss:prior}) is the prior distribution of $\phi$ for a given model $z$ and can be denoted by $p( \phi \mid z)$. 

Now, under the given model $z$, the posterior distribution in (\ref{post1}) can be written as
\begin{eqnarray}
p(\phi, \theta\mid  \mbox{data}, z)\propto g\left\lbrace S_n(\phi), U_{PS}(\theta,\phi)\mid \theta, \phi\right\rbrace p(\phi \mid z )p(\theta).
\label{post2}
\end{eqnarray}
Note that, in (\ref{post2}), we can express the joint density as a product of the marginal density of $S_n(\phi)$ and the conditional density of $U_{PS} ( \theta, \phi)$ given $S_n(\phi)$. That is,
\begin{equation}
 g\left\lbrace S_n(\phi), U_{PS}(\theta,\phi)\mid \theta, \phi\right\rbrace  = g_1 \left\lbrace S_n(\phi)\mid  \phi\right\rbrace g_2\left\lbrace   U_{PS}(\theta,\phi)\mid   S_n(\phi),  \theta, \phi\right\rbrace,\label{post0}
\end{equation}
where $g_1(\cdot), g_2(\cdot)$ are the density functions derived from the joint asymptotic  normality in (\ref{claim1}).

Thus, combining (\ref{post2}) with (\ref{post0}), the posterior distribution in (\ref{post2}) can be written as 
\begin{equation}
 p(\phi, \theta \mid \mbox{data},  z ) = p_1 ( \phi \mid \mbox{data},  z ) p_2 ( \theta \mid \mbox{data}, \phi) ,
 \label{post3}
\end{equation}
where
\begin{equation}
  p_1 ( \phi \mid \mbox{data},  z ) = \frac{  g_1 \left\lbrace S_n(\phi)\mid \phi\right\rbrace p(\phi \mid z) }{ \int g_1 \left\lbrace S_n(\phi)\mid \phi\right\rbrace p(\phi \mid z) d \phi}
\label{step2}
\end{equation}
and
\begin{equation}
p_2 ( \theta \mid \mbox{data}, \phi) = \frac{ g_2\left\lbrace   U_{PS}(\theta,\phi)\mid   S_n(\phi),  \theta, \phi\right\rbrace p( \theta ) }{ \int g_2\left\lbrace   U_{PS}(\theta,\phi)\mid   S_n(\phi),  \theta, \phi\right\rbrace p( \theta ) d \theta } .
 \label{step3}
\end{equation}


Therefore, following the standard Bayesian procedure,  the  posterior distribution of $(\phi, \theta, z)$ can be obtained from
\begin{eqnarray}
&p(\phi, \theta,z \mid \mbox{data} )&=\frac{ p(\phi, \theta\mid  \mbox{data}, z) p(z) }{\int \int \int  p(\phi, \theta\mid  \mbox{data}, z) p(z) dz d\phi d\theta}\nonumber\\
&&=\frac{p_1(\phi\mid\mbox{data}, z)p_2(\theta\mid \mbox{data}, \phi)p(z)}{\int \int \int p_1(\phi\mid\mbox{data}, z)p_2(\theta\mid \mbox{data}, \phi)p(z) dz d\phi d \theta},\label{Joint_P}
\end{eqnarray}
where $p(z)$ is the prior distribution of $z$  in (\ref{ber:prior}), $p_1(\phi\mid \mbox{data},z)$ is the posterior distribution of $\phi$ in (\ref{step2}), and $p_2(\theta\mid  \mbox{data}, \phi)$ is the posterior distribution of $\theta$ in (\ref{step3}).  

Using the Gibbs sampling \citep{casella1992explaining} procedures, our proposed Bayesian sparse propensity score (BSPS) method can be described by the following two steps:
\begin{itemize}
	\item []\textit{Step 1 (Model step)}: Given $(\phi^{(t)}, \theta^{(t)})$, generate model $z^{(t+1)}$ from $p(z\mid  \mbox{data}; \phi^{(t)}, \theta^{(t)})$.
	\item[]\textit{Step 2 (Posterior step)}: Given $z^{(t+1)}$, generate $(\phi^{(t+1)}, \theta^{(t+1)})$ from $p(\phi, \theta\mid \mbox{data}, z^{(t+1)})$.
\end{itemize}
\textit{Step 1} is the new step for model selection. \textit{Step 2} is already discussed in \cite{BayesianPS}.

We first discuss \textit{Step 1}. Using (\ref{post3}), the posterior distribution of $z$ given $(\phi^{(t)}, \theta^{(t)})$ can be derived as
\begin{eqnarray*}
p(z\mid \mbox{data}; \phi^{(t)}, \theta^{(t)})&=&\frac{p(\phi^{(t)},\theta^{(t)}\mid \mbox{data},z) p(z)}{\int p(\phi^{(t)},\theta^{(t)}\mid \mbox{data},z) p(z) dz}\\
&=&\frac{L( \phi^{(t)}\mid \mbox{data} ) p( \phi^{(t)} \mid z)  p(z)}{\int  L( \phi^{(t)} \mid \mbox{data})    p(\phi^{(t)}\mid z)p(z) dz},\nonumber\\
&=&\frac{p(\phi^{(t)}\mid z)p(z)}{\int p(\phi^{(t)}\mid z)p(z) d z},\nonumber
\end{eqnarray*}
where  $L(\phi\mid\mbox{data})=\exp\left\lbrace l_n(\phi)\right\rbrace $ is the likelihood function of $\phi$. 
 Thus, using (\ref{ss:prior}) and (\ref{ber:prior}),
\textit{Step 1} can be simplified as generating $z^{(t+1)}=(z_1^{(t+1)},z_2^{(t+1)},\ldots,z_p^{(t+1)})^\T$ from
\begin{eqnarray}
z^{(t+1)}_j\overset{ind}{\sim}\text{Ber}\left(\frac{w_j \psi(\phi^{(t)}_j|0,\nu_1)}{w_j \psi(\phi^{(t)}_j|0,\nu_1)+(1-w_j)\psi(\phi^{(t)}_j|0,\nu_0)} \right), ~j=1,2,\ldots,p,\label{step 1}
\end{eqnarray}
where $\psi(\cdot|\mu,\sigma^2)$ denotes a Gaussian density function with mean $\mu$ and variance $\sigma^2$. Thus, \textit{Step 1} does not require any iterative algorithm and hence computationally efficient.

For \textit{Step 2}, given $z^{(t+1)}$, we can use (\ref{post3}) to generate the posterior values by the following two steps:
\begin{itemize}
		\item[] \textit{Step 2a}: Given  $z^{(t+1)}$, generate $ \phi^{(t+1)}$ from $p_1(\phi\mid \mbox{data}, z^{(t+1)})$ in (\ref{step2}).
	\item[] \textit{Step 2b}: Given $\phi^{(t+1)}$, generate $\theta^{(t+1)}$ from $p_2( \theta \mid \mbox{data}, \phi^{(t+1)} )$ in (\ref{step3}).
\end{itemize}
For \textit{Step 2a}, since the likelihood of $\phi$ is known, we can use
\begin{eqnarray}
p(\phi\mid \mbox{data}, z^{(t+1)})=\frac{L(\phi\mid\mbox{data})p(\phi\mid z^{(t+1)})}{\int L(\phi\mid \mbox{data})p(\phi\mid z^{(t+1)}) d\phi}\label{post_phi1}
\end{eqnarray}
to generate the posterior of $\phi$ given the model $z^{(t+1)}$ and data. 
In \textit{Step 2b},  $\theta^{(t+1)}$ are generated from $p_2( \theta \mid \mbox{data},  \phi^{(t+1)} )$ in (\ref{step3}), where the conditional distribution is derived from the joint normality in (\ref{claim1}). 
The computational details of generating the posterior values from \textit{Step 2} efficiently are described in Appendix \ref{App::A}. 

\section{ Asymptotic properties}\label{sec::asymp}
To establish the asymptotic properties, we first assume the regularity conditions for the existence of the unique solution to $S_n(\phi)=0$,
as discussed in \cite{silvapulle1981existence}.
To establish the asymptotic properties of the PS estimator under a high dimensional setup, assume $X$ can be decomposed as $X=(X_1, X_2, X_3)$, where $(X_1,X_2)$ satisfy $\mathrm{P}(\delta=1\mid X)=\mathrm{P}(\delta=1\mid X_1)$ and $\mathrm{P}(Y\mid X)=\mathrm{P}(Y\mid X_1, X_2)$. Note that $X_3$ is not helpful in explaining $(\delta, Y)$.
 Let $p_1,p_2,p_3$ be the dimension of $X_1, X_2, X_3$, respectively, such that $p=p_1+p_2+p_3$. Let $U_n(\eta)=\left\lbrace S_n^\T(\phi),U_{PS}^\T(\theta,\phi)\right\rbrace^\T$ and $\eta=(\phi, \theta)$. We now make the following assumption:

\begin{itemize}
	\item[]\textbf{(A1)} In a neighborhood of the true parameters $\eta_0=(\phi_0, \theta_0)$,  assume $E\left\lbrace U_n(\eta_0)\right\rbrace=0 $, $E\left\lbrace \left|\partial U_n(\eta)/ \partial \eta_j\right| \right\rbrace <\infty$ and $E\left\lbrace\left| \partial^2 U_n(\eta)/ \partial \eta^T \partial \eta_j \right|\right\rbrace <\infty$ hold. 
\end{itemize}
 Condition (\textbf{A1}) is the usual regularity conditions for PS estimation. 
Define $\hat \theta_{PS}(X)$ to be the PS estimator of $\theta$ using the covariate $X$ for the PS model and 
$\hat \pi_i=G(X_i^\T\hat \phi)$ for computing the PS estimator. Let $\hat \theta_{PS}=\hat \theta_{PS}(X_1, X_2, X_3)$ for simplicity.

\begin{theorem}\label{high_PS} 
	Assume that the solution to (\ref{sec2:eq6}) is unique.
	Under assumptions (\textbf{A1}) and MAR assumption in (\ref{sec2:eq2}), we can establish the following.
	\begin{itemize}
		\item [1.] The bias of the PS estimator satisfies
		\begin{eqnarray}
		E\left( \hat \theta_{PS}-\theta_0\right) =O\left( \frac{p}{n} \right) \label{bias}.
		\end{eqnarray}
		\item [2.] The variance of the PS estimator  has
		\begin{eqnarray}
		\mathrm{var}\left( \hat \theta_{PS}-\theta_0\right) =O\left\lbrace \max \left(\frac{1}{n},\frac{p}{n^2}+\frac{p^2}{n^3} \right) \right\rbrace.\label{variance}
		\end{eqnarray}
		\item[3.] The PS estimator including  $(X_1, X_2)$ is more efficient than  the PS estimator including $X_1$ only, in the sense of
		\begin{eqnarray}
		\mathrm{Var}\left\lbrace  \hat \theta_{PS}(X_1, X_2)\right\rbrace \leq 	\mathrm{Var}\left\lbrace  \hat \theta_{PS}(X_1)\right\rbrace.\nonumber
		\end{eqnarray}
	\end{itemize}
\end{theorem}
The proof of Theorem \ref{high_PS} is shown in Appendix \ref{proof_theorem1}. In (\ref{bias}), the bias of the PS estimator depends on the order of $p$. If $p$ is bounded, then the bias of PS estimator is asymptotically negligible. If $p_3$ increases with $n$, then the PS estimator can be significant biased. From the first two statements of Theorem \ref{high_PS}, we can see that the PS estimator is significantly biased and its variance keeps increasing as $p_3$ increases. Under the sparsity setup, the true response model is not necessarily optimal. That is, including $X_2$ which is correlated with $Y$ helps to improve the performance of the PS estimator.

We now establish the model selection consistency under the Bayesian framework. The Bayesian model selection consistency
is satisfied if the posterior probability of the true model converges
to one as the sample size goes to infinity \citep{Casella:2009}. To achieve the model selection consistency or Oracle property \citep{Fan:2001,Zou:2006}, we further assume the following condition.
\begin{itemize}
	\item[](\textbf{A2}) Assume $p_1=O(1)$ and $p_2=O(1)$. 
	\item [](\textbf{A3}) In the Spike-Slab prior in (\ref{ss:prior}), $\nu_0=o(n^{-1})$, $\nu_1=O(n)$, and $w_1=w_2=\cdots=w_p=0.5$.\label{asump:a6}
\end{itemize}
Condition (\textbf{A1}) is the sparsity assumption.
The choice of $w_j=0.5$ represents a non-informative prior for each covariate component. The following theorem establishes the Oracle property of the proposed Bayesian sparse propensity score method.

\begin{theorem}\label{thm:model_consistnecy}
	Under assumptions (\textbf{A1})--(\textbf{A3}), $p_3=o(n)$ and the MAR assumption in (\ref{sec2:eq2}), we have
	\begin{eqnarray*}
		p(z=z_o|\mbox{data})\xrightarrow{} 1,
	\end{eqnarray*}
	in probability, where $z_o$ is the true response model and $	p(z|\mbox{data})$ is the marginal posterior probability in (\ref{step 1}).

\end{theorem}

The proof of Theorem \ref{thm:model_consistnecy} is given in Appendix \ref{proof_theorem2}. According to Theorem \ref{thm:model_consistnecy}, we  observe that the probability that \textit{Step 1} selects the true model becomes very close to one when the sample size $n$ is sufficiently large.
Thus, the proposed Bayesian method can effectively eliminate irrelevant covariates and select important ones to adjust for nonresponse bias.

	Note that, in Theorem \ref{thm:model_consistnecy}, we assume $p_3=o(n)$, which can be extended with a small modification in \textit{Step 1}. Instead of using non-informative priors, we use this assumption to make the prior distribution to satisfy $\mathrm{P}\left\lbrace \sum_{j}z_j=o(n)\right\rbrace =1$. That can be implemented  as dropping the generated candidate models until we obtain $\sum_{j}z_j=o(n)$. Thus, our proposed method can be easily extended to $p_3=O(n)$ and the ultra-high dimensional setup of \cite{chen2008extended}.

 Since we assume the true response model is sparse,  $p_o=\sum_jz_{o,j}$ is fixed with increasing $n$. Thus, the asymptotic normality can be established under the regularity conditions.
\begin{theorem}\label{cor:1}
	Under the conditions in Theorem \ref{thm:model_consistnecy} and the regularity conditions of \citet{BayesianPS}, we have
	\begin{eqnarray}
	\left\lbrace \hat{V}ar(\hat \theta_{\text{BSPS}})\right\rbrace^{-1/2} \left(  \hat \theta_{\text{BSPS}}-\theta_0\right) \xrightarrow{d} \text{N}\left(0, 1 \right),\nonumber
	\end{eqnarray}
	where $\hat \theta_{\text{BSPS}}=M^{-1} \sum_{k=1}^{M}\theta^*_{(k)}$ and $\theta^*_{(k)}$ are generated from (\ref{Joint_P}), and $	\hat{V}ar(\hat \theta_{\text{BSPS}})=\sum_{k=1}^{M}\left(\theta^*_{(k)}- \hat \theta_{\text{BSPS}}\right)^2/(M-1)$.
\end{theorem}
\citet{BayesianPS} have already established the asymptotic normality of the Bayesian propensity score (BPS) estimator under the correctly specified response model. By Theorem \ref{thm:model_consistnecy}, the probability that \textit{Step 1} selects the true model converges to one. Consequently, the asymptotic distribution of our BSPS estimator is the same as the asymptotic distribution of BPS estimator under the true model, which leads to the asymptotic normality of the BSPS estimator.

\begin{remark}
	From Theorem \ref{thm:model_consistnecy}, we can see that the model uncertainty of $z$ vanishes as $n\xrightarrow{}\infty$. However, in finite samples, model selection always contributes to the variability of $\hat\theta_{\text{BSPS}}$. The advantage of the proposed Bayesian method lies in its capture of the variability of the model uncertainty in the finite sample case.  The posterior distribution of $z$ in \textit{Step 1} incorporates the model selection uncertainty automatically.
By the Law of Large Numbers (LLN), we can show that
	\begin{eqnarray}
	&(M-1)^{-1}\sum_{k=1}^{M}\left(\theta^*_{(k)}- \hat \theta_{\text{BSPS}}\right)^2&\xrightarrow{P}Var\left\lbrace \theta^*\mid \mbox{data}\right\rbrace \nonumber\\
	&&=Var\left\lbrace  E\left(  \theta^*\mid z^*,\mbox{data}\right)  \mid \mbox{data}\right\rbrace +E\left\lbrace  Var\left( \theta^*\mid z^*,\mbox{data}\right)\mid \mbox{data} \right\rbrace  \nonumber,
	\end{eqnarray}
	where $\theta^*$ is generated from \textit{Step 2} given model $z^*$. In the finite sample, $Var\left\lbrace  E\left(  \theta^*\mid z^*,\mbox{data}\right) \mid \mbox{data} \right\rbrace$ represents the variability due to the model uncertainty.  Bt Theorem \ref{thm:model_consistnecy},  $Pr(z^*=z_o\mid \mbox{data})=1$, as $n\xrightarrow{}\infty$, which leads to $Var\left\lbrace  E\left(  \theta^*\mid z^*,\mbox{data}\right)  \mid \mbox{data}\right\rbrace=0$.
\end{remark}

\section{Optimal Bayesian sparse propensity score estimation}\label{sec::extension}
In Section \ref{sec::asymp}, Theorem \ref{high_PS} points to the efficiency gain in including the covariates $(X_2)$, which are correlated with the outcome variable, into the PS model. However, the sparse Bayesian method in Section \ref{sec:proposed} only selects the covariates involved in the true response model.
Therefore, in this section, we propose an optimal Bayesian sparse propensity score estimation that improves the efficiency by incorporating the relevant auxiliary variables from the full sample.

To select the covariates correlated with $Y$ given that $X_1$ is selected, we propose to use the following ``working" outcome model 
\begin{eqnarray}
y_i=x_i^\T\beta+e_i,\label{working}
\end{eqnarray}
where $\beta=(\beta_1, \beta_2, \cdots, \beta_p)^\T$ and $e_i\sim N(0, \sigma_e^2)$ independently. Let $f_w(y_i\mid x_i)$ be the density function for the working model in (\ref{working}).
Note that, the purpose of the outcome model in (\ref{working}) is to select important covariates in addition to $X_1$ to improve efficiency. The validity of the resulting PS estimator does not require the outcome model assumption (\ref{working}) to hold. Therefore, the same model selection method using the Spike-and-Slab prior can also be used.  Let $u=(u_1, u_2, \cdots, u_p)^T$, where
\begin{eqnarray}
u_j=\left\lbrace
\begin{array}{lr}
1 & \text{if $\beta_j\neq 0$}\\
0 & \text{otherwise}.\nonumber
\end{array}\right.
\end{eqnarray}
To select additional relevant variables given $X_1$, we can assign 
\begin{eqnarray}
\beta_j|u_j&\overset{ind}{\sim}& \text{N}(0,\gamma_{0}(1-u_j)+\gamma_{1}u_j),\nonumber\\
u_j\mid z_j&\overset{ind}{\sim} &\text{Ber}\left\lbrace z_j+(1-z_j)\xi_j\right\rbrace ,\label{uz}
\end{eqnarray}
where $\xi_j\in (0,1)$ and $(\gamma_0,\gamma_1)$ are fixed to be small and large values, respectively. The prior distribution for $u_i$ is informative for $z_i=1$, as we want to include $X_1$ in advance before including $X_2$.

Let the prior distribution of $\sigma_e^2$, say $p(\sigma_e^2)$, be the inverse gamma distribution with parameters $(c_1,c_2)$. We set $c_1=c_2=10^{-7}$ to create a non-informative prior. Then, the posterior distribution
\begin{eqnarray}
p(\beta, u, \sigma_e^2\mid \mbox{data},z)\propto L_w(\beta, \sigma_e^2) p(\beta\mid u) p(u\mid z) p(\sigma_e^2),\label{post_u}
\end{eqnarray}
where $ L_w(\beta, \sigma_e^2)=\prod_{\delta_i=1} f_w(y_i\mid x_i; \beta, \sigma_e^2)$.
To generate $u$ from  the posterior distribution, the same data augmentation method \citep{tanner1987calculation} can be used. The implementation of (\ref{post_u}) can be described as follows.
\begin{itemize}
	\item []\textit{I-step}: Given $\beta^{(t)}$ and $\sigma_e^{2(t)}$, generate $u^{(t+1)}$ from
	\begin{eqnarray}
	u_j^{(t+1)}\overset{ind}{\sim} \text{Ber}\left( z_j+(1-z_j)\frac{\xi_j \psi(\beta_j^{(t)}\mid 0, \gamma_1)}{\xi_j \psi(\beta_j^{(t)}\mid 0, \gamma_1)+(1-w_j)w_j \psi(\beta_j^{(t)}\mid 0, \gamma_0)}\right), \label{I-step}
	\end{eqnarray}
	for $j=1,2,\cdots, p.$
	\item[]\textit{P-step}: Given $u^{(t+1)}$, generate $(\beta^{(t+1)}, \sigma_e^{2(t+1)})$ from 
	\begin{eqnarray}
	p(\beta, \sigma_e^2\mid \mbox{data}, u^{(t+1)})=\frac{L_w(\beta, \sigma_e^2)p(\beta\mid u^{(t+1)})p(\sigma_e^2)}{\int L_w(\beta, \sigma_e^2)p(\beta\mid u^{(t+1)})p(\sigma_e^2) d\beta d\sigma_e^2}.\label{pstep}
	\end{eqnarray}
	The detailed algorithm for generating $(\beta, \sigma_e^2)$ from (\ref{pstep}) is described in Appendix \ref{App::B}.
\end{itemize}


 Given the augmented PS model  $u^*$, the following reduced estimating equations can be used to estimate the optimal $\theta$ without any model assumptions on $Y$.
 \begin{eqnarray}
 U_{opt}(\phi_{u^*}, \theta)=\left\lbrace
 \begin{array}{l}
\sum_{i=1}^n \frac{\delta_i}{\pi(\phi_{u^*}; x_{i, u^*})} (y_i-\theta)\\
\sum_{i=1}^n \left\lbrace \delta_i -\pi(\phi_{u^*}; x_{i, u^*})\right\rbrace x_{i, u^*}\\
\sum_{i=1}^n\left\lbrace \frac{\delta_i}{\pi(\phi_{u^*}; x_{i, u^*})}-1 \right\rbrace x_{i, u^*}
 \end{array}\right\rbrace =0,\nonumber
 \end{eqnarray}
 where $\phi_{u^*}$ and $x_{i, u^*}$ are respectively sub-vectors of $\phi$ and $x_i$ corresponding to the chosen model $u^*$. Given $u^*$, let $\zeta^*=(\phi_{u^*}, \theta)$.
To generate the posterior distribution of $\theta$ given $u^*$ and  $ U_{opt}(\zeta^*)$, approximate Bayesian computation can be used. Under some regularity conditions, we can establish that
\begin{eqnarray}
 n^{-1/2}U_{opt}(\zeta^*)\mid \zeta^*, u^* \xrightarrow{} N\left\lbrace 0, \Sigma_{opt}(\zeta^*)\right\rbrace. \label{asym3}
\end{eqnarray}
Using the asymptotic sampling distribution in (\ref{asym3}) to replace the role of likelihood, the posterior distribution of $\zeta^*$ can be generated from
\begin{eqnarray}
p\left\lbrace \zeta^*\mid u^*, U(\zeta^*)\right\rbrace =\frac{g\left\lbrace U_{opt}(\zeta^*); \zeta^*, u^* \right\rbrace p(\zeta^*)}{\int g\left\lbrace U_{opt}(\zeta^*); \zeta^*, u^* \right\rbrace p(\zeta^*) \mathrm{d}\zeta^*}, \label{pos_zeta}
\end{eqnarray}
where $g\left\lbrace U_{opt}(\zeta^*); \zeta^*, u^* \right\rbrace$ is the density function from (\ref{asym3}) and $p(\zeta^*)\propto 1$ is a flat prior.

In summary, our proposed optimal Bayesian sparse propensity score (OBSPS) method can be described as follows.
\begin{itemize}
	\item [S1.] Use \textit{Step 1} and \textit{Step 2} in Section \ref{sec:proposed} to generate $z^*$.
	\item[S2.] Given $z^*$, use \textit{I-step} and \textit{P-step} in (\ref{I-step}) and (\ref{pstep}) to generate $u^*$.
	\item[S3.]  Given $u^*$, generate the posterior distribution of $\theta$ from (\ref{pos_zeta}).
\end{itemize}


\section{Simulation studies}\label{sec::simulation}

In this section, we conduct two simulation studies to examine the finite sample performance of the proposed Bayesian method. The first simulation study investigates the proposed Bayesian method under the IID setup. In the second simulation study, we apply our proposed method using real data obtained from the 2006 Korean Labor and Income Panel Study (KLIPS).

\subsection{ Simulation study I}

In the first simulation, our data generation process consists of the following two parts.
\begin{itemize}
	\item[1.] Generate a random sample of size $n=200$, $\{(x_i,y_i):i=1,2,\ldots,n\}$, from each of the following models:
	\begin{eqnarray}
	\mathcal{M}_1&:&y_i\overset{ind}{\sim}2x_{i1}+2x_{i3}+e_{i};\nonumber\\
	\mathcal{M}_2&:&y_i\overset{ind}{\sim} 1.5x_{i1}+0.5x^2_{i3} +2x_{i4}+e_i ;\nonumber
	\end{eqnarray}
	where  $x_i=(x_{i1},x_{i2},\ldots,x_{i,p+1})^\T$ with $x_{i1}=1$, and the errors $e_i$ are generated independently from $N(0,1)$.
The covariates $x_{i2},x_{i3},\ldots,x_{ip} $ are independently generated from $N(0, S)$, where $S=\left(\rho^{|i-j|}\right)_{1\leq i,j\leq p}$.   We use two values for $\rho$: $\rho=0$ for independent covariates and $\rho=0.5$ for correlated covariates.
Also, we use $p=10,50,$ and $100$. 
	\item[2.] For $i=1,2,\ldots,n$, generate the response indicator of $y_i$ from the following response mechanism:
	\begin{eqnarray}
\delta_i\overset{ind}{\sim} \text{Bernoulli}\left\{\frac{\exp(x_{i1}+x_{i2})}{1+\exp(x_{i1}+x_{i2})}\right\};\nonumber
	\end{eqnarray}
\end{itemize}
 Note that in our setup $p$ controls the amount of sparsity on the propensity score. As $p$ increases, the propensity score becomes more sparse. We are interested in estimating $\theta =E(Y)$.

For each setup, we generate $B=2,000$ Monte Carlo samples of size $n=200$ and  we apply the following methods:

\begin{itemize}
	\item [1.] PS: The traditional PS estimator, say $(\hat{\phi}_\text{PS}, \hat{\theta}_\text{PS})$, is obtained by solving the joint estimating equations
	\begin{eqnarray*}
		\frac{1}{n}\sum_{i=1}^{n}\left\lbrace \delta_i-\pi(x_i; \phi)\right\rbrace x_i=0,\\
		\frac{1}{n}\sum_{i=1}^{n}\frac{\delta_i}{\pi(x_i;\phi)}(y_i-\theta)=0,
	\end{eqnarray*}
	where $\pi(x_i;\phi)=G(x_i^T\phi)$.
	The variance of $(\hat{\phi}_\text{PS}, \hat{\theta}_\text{PS})$ is estimated by Taylor linearization. The 95\% confidence intervals are constructed from the asymptotic normal distribution of $(\hat{\phi}_\text{PS}, \hat{\theta}_\text{PS})$.
	\item [2.] TPS:  The true propensity score (TPS) method in which the ordinary PS method is applied using the covariates in the true response mechanism. The 95\% confidence intervals are constructed from the asymptotic normal distribution of $(\hat{\phi}_\text{TPS}, \hat{\theta}_\text{TPS})$
	\item [3.] LASSO: We first apply the LASSO method to select the response model with $\lambda$ in (\ref{lasso}) chosen by the default 5-fold cross-validation method. The algorithm is implemented in ``glmnet'' \citep{friedman2009glmnet}.  Then we apply the traditional PS estimation method to the selected response model. Variances and confidence intervals are obtained by using the asymptotic normal distribution of $(\hat{\phi}_\text{LASSO}, \hat{\theta}_\text{LASSO})$ for the selected response model.
	\item [4.] BSPS: The Bayesian sparse PS method proposed in Section \ref{sec:proposed}.  In BSPS, we set $w_1=\cdots=w_p=0.5$, $\nu_0=10^{-4}$, and $\nu_1=10^4$ to induce noninformative priors. Using the formula in Section \ref{sec:proposed}, we compute the BSPS estimate and its variance estimate based on the posterior sample of size $2,000$ after $2,000$ burn-in iterations. The 95\% confidence intervals are constructed from the quantiles of the posterior sample.
	\item[5.] OBSPS: The optimal Bayesian sparse PS method proposed in Section \ref{sec::extension}. In OBSPS, we use the same setup in BSPS and let $\xi_1=\cdots=\xi_p=0.5, \gamma_0=10^{-4}, \gamma_1=10^4$.
\end{itemize}

To assess the variable selection performance of BSPS, OBSPS, and LASSO methods, we compute true positive rate (TPR) and true negative rate (TNR), where TPR is the proportion of the regression coefficients that are correctly identified as nonzero and TNR is the proportion of the regression coefficients that are correctly identified as zero. The coverage probabilities of each methods are computed by counting how often the confidence intervals contain the true parameter values. In BSPS and LASSO, we present TPR and TNR for the response model. In OBSPS, we show TPR and TNR for the working model to select correlated covariates.
 The simulation results for models $\mathcal{M}_1$ and $\mathcal{M}_2$ are presented in Tables \ref{tbl:sbps} and \ref{tbl:sbps_res2_model1}, respectively.

\begin{table}[ht]
	\centering
	\caption{Simulation results for $\mathcal{M}_1$:  ``Bias" is the bias of the point estimator for $\theta$, ``S.E." represents the standard error of the point estimator, ``$E[\text{S.E.}]$" is the average of the estimated standard error, ``CP" represents the coverage probability of the 95\% confidence interval estimate.
	}\label{tbl:sbps}
	\vskip .2cm
	\centerline{\tabcolsep=3truept
		\centering
	\begin{tabular}{l|l|l|rrrr|rr}
		\hline
		$\rho$  &p & Method & Rbias$\times 100$ & S.E.$\times 100$  & $E[\text{S.E.}]$$\times 100$ & CP$\times 100$  & TPR & NPR \\
		\hline
	\multirow{15}{*}{0} & \multirow{5}{*}{10}& PS & -0.6 & 3.4 & 2.7 & 93.2 &  &  \\
	&  & TPS & -0.6 & 3.2 & 2.8 & 94.1 &  &  \\
	&  & LASSO & -0.8 & 3.7 & 3.6 & 95.8 & 1.0 & 0.8 \\
	&  & BSPS & -0.6 & 4.2 & 3.8 & 94.6 & 1.0 & 1.0 \\
	&  & OBSPS & -0.4 & 3.0 & 2.7 & 93.3 & 1.0 & 1.0 \\
	\cline{2-9}
	&  \multirow{5}{*}{50}& PS & -0.6 & 9.4 & 1.8 & 78.7 &  &  \\
	&  & TPS & -0.2 & 3.0 & 2.8 & 94.5 &  &  \\
	&  & LASSO & -0.2 & 4.5 & 3.8 & 94.5 & 1.0 & 0.9 \\
	&  & BSPS & -0.4 & 4.1 & 3.8 & 94.3 & 1.0 & 1.0 \\
	&  & OBSPS & 0.0 & 2.9 & 2.7 & 94.0 & 1.0 & 1.0 \\
	\cline{2-9}
	&  \multirow{5}{*}{100}& PS &  &  &  &  &  &  \\
	&  & TPS & 0.4 & 3.2 & 2.8 & 94.4 &  &  \\
	&  & LASSO & 0.4 & 5.6 & 3.9 & 92.1 & 1.0 & 0.9 \\
	&  & BSPS & 0.4 & 4.2 & 3.9 & 94.7 & 1.0 & 1.0 \\
	&  & OBSPS & 0.2 & 3.0 & 2.7 & 94.2 & 1.0 & 1.0 \\
\hline
\multirow{15}{*}{0.5} & \multirow{5}{*}{10} & PS & 0.1 & 3.6 & 2.7 & 92.9 &  &  \\
	&  & TPS & 0.0 & 3.2 & 2.9 & 94.2 &  &  \\
	&  & LASSO & -2.2 & 3.7 & 3.4 & 94.6 & 1.0 & 0.8 \\
	&  & BSPS & -0.0 & 4.0 & 3.6 & 93.7 & 1.0 & 1.0 \\
	&  & OBSPS & 2.0 & 2.9 & 2.7 & 93.5 & 1.0 & 1.0 \\
	\cline{2-9}
	&  \multirow{5}{*}{50} & PS & 0.2 & 10.0 & 1.8 & 76.7 &  &  \\
	&  & TPS & 0.0 & 3.3 & 2.8 & 93.9 &  &  \\
	&  & LASSO & -1.8 & 4.4 & 3.5 & 92.9 & 1.0 & 0.9 \\
	&  & BSPS & 0.0 & 4.1 & 3.7 & 93.6 & 1.0 & 1.0 \\
	&  & OBSPS & 2.4 & 3.0 & 2.7 & 93.3 & 1.0 & 1.0 \\
	\cline{2-9}
	&  \multirow{5}{*}{100} & PS &  &  &  &  &  &  \\
	&  & TPS & 0.4 & 3.1 & 2.8 & 94.4 &  &  \\
	&  & LASSO & -0.0 & 5.6 & 3.8 & 93.0 & 1.0 & 0.9 \\
	&  & BSPS & -0.0 & 4.1 & 3.9 & 95.3 & 1.0 & 1.0 \\
	&  & OBSPS & 0.2 & 3.0 & 2.7 & 94.0 & 1.0 & 1.0 \\
	\hline
	\end{tabular}
	}
\end{table}

\begin{table}[ht]
	\centering
	\caption{Simulation results for $\mathcal{M}_2$:  ``RBias" is the relative bias of the point estimator for $\theta$, ``S.E." represents the standard error of the point estimator, ``$E[\text{S.E.}]$" is the average of the estimated standard error, ``CP" represents the coverage probability of the 95\% confidence interval estimate}. \label{tbl:sbps_res2_model1}
	\vskip .2cm
	\centerline{\tabcolsep=3truept
	\begin{tabular}{l|l|l|rrrr|rr}
	\hline
	$\rho$  &p & Method & Rbias$\times 100$ & S.E.$\times 100$  & $E[\text{S.E.}]$$\times 100$ & CP$\times 100$  & TPR & NPR \\
	\hline
\multirow{15}{*}{0} & \multirow{5}{*}{10}  & PS & 0.2 & 3.7 & 3.1 & 92.7 &  &  \\
&  & TPS & 0.4 & 4.5 & 4.2 & 94.2 &  &  \\
&  & LASSO & 0.4 & 4.0 & 4.1 & 95.1 & 1.0 & 0.8 \\
&  & BSPS & 0.4 & 4.4 & 4.2 & 94.0 & 1.0 & 1.0 \\
&  & OBSPS & 0.2 & 3.4 & 3.1 & 93.5 & 1.0 & 1.0 \\
\cline{2-9}
&  \multirow{5}{*}{50} & PS & 0.0 & 9.5 & 2.1 & 80.2 &  &  \\
&  & TPS & 0.2 & 4.5 & 4.2 & 94.8 &  &  \\
&  & LASSO & 0.2 & 5.0 & 4.2 & 94.2 & 1.0 & 0.9 \\
&  & BSPS & 0.0 & 4.4 & 4.2 & 95.0 & 1.0 & 1.0 \\
&  & OBSPS & -0.0 & 3.2 & 3.1 & 94.3 & 1.0 & 1.0 \\
\cline{2-9}
&  \multirow{5}{*}{100} & PS &  &  &  &  &  &  \\
&  & TPS & -0.2 & 4.7 & 4.2 & 93.9 &  &  \\
&  & LASSO & -0.2 & 6.3 & 4.2 & 91.9 & 1.0 & 0.9 \\
&  & BSPS & -0.2 & 4.6 & 4.2 & 93.6 & 1.0 & 1.0 \\
&  & OBSPS & -0.4 & 3.4 & 3.1 & 93.5 & 1.0 & 1.0 \\
\hline
\multirow{15}{*}{0.5} & \multirow{5}{*}{10} & PS & 0.8 & 3.7 & 3.1 & 92.7 &  &  \\
&  & TPS & 0.8 & 4.2 & 3.9 & 93.7 &  &  \\
&  & LASSO & -0.8 & 3.9 & 3.8 & 94.3 & 1.0 & 0.8 \\
&  & BSPS & 0.8 & 4.3 & 4.0 & 93.9 & 1.0 & 1.0 \\
&  & OBSPS & 1.0 & 3.4 & 3.1 & 93.4 & 1.0 & 1.0 \\
\cline{2-9}
&   \multirow{5}{*}{50} & PS & -0.4 & 7.3 & 2.1 & 80.9 &  &  \\
&  & TPS & -0.4 & 4.2 & 3.9 & 94.6 &  &  \\
&  & LASSO & -2.0 & 4.5 & 4.0 & 94.2 & 1.0 & 0.9 \\
&  & BSPS & -0.4 & 4.3 & 4.0 & 94.7 & 1.0 & 1.0 \\
&  & OBSPS & -0.2 & 3.3 & 3.1 & 94.1 & 1.0 & 1.0 \\
\cline{2-9}
&  \multirow{5}{*}{100}  & PS &  &  &  &  &  &  \\
&  & TPS & -0.2 & 4.2 & 3.9 & 94.3 &  &  \\
&  & LASSO & -1.6 & 5.3 & 4.0 & 93.0 & 1.0 & 0.9 \\
&  & BSPS & -0.0 & 4.2 & 4.0 & 94.7 & 1.0 & 1.0 \\
&  & OBSPS & -0.0 & 3.3 & 3.1 & 93.9 & 1.0 & 1.0 \\
	\hline
\end{tabular}
}
\end{table}

Table \ref{tbl:sbps} shows the numerical results for $\mathcal{M}_1$.  Overall, the proposed methods perform similarly between correlated covariates $(\rho=0.5)$ and independent covariates $(\rho=0)$.  When dimension is low, specifically, when p=10, PS and TPS  have similar performance in terms of bias and standard errors. TPS are more efficient than PS due to sparsity. LASSO and BSPS can select the true response model with large probabilities. However, LASSO and BSPS obtain larger standard errors than TPS due to additional model uncertainty under finite samples. Overall, BSPS outperforms LASSO in term of model consistency. OBSPS always provides the most efficient estimators by incorporating relevant auxiliary variables from the full sample. For small $p$, all methods achieve approximately 95\% coverage probabilities for corresponding confidence  intervals or credible intervals for Bayesian models.

When $p$ increases to $50$ in $\mathcal{M}_1$, the PS estimator using all variables shows large standard errors. Moreover, the average of the estimated standard errors for the PS estimator is much smaller than the true standard error of the PS estimator, which leads to biased interval estimation and low coverage probability.  Note that TPS is the gold standard method, where we pretend that we know the truth. Thus, TPS is invariant for large $p$. BSPS performs better than LASSO for selecting the true response model. Therefore, the model uncertainty of BSPS and OBSPS are much smaller than LASSO. Table \ref{tbl:sbps} shows that Monte Carlo standard error of LASSO is much larger than estimated standard error due to model uncertainty. However, the increased variances of BSPS and OBSPS are not as  obvious as LASSO due to better model selection performance. In summary, BSPS obtains comparable estimator and inference with TPS. OBSPS is still most efficient relative to all other methods and it identifies the relevant covariates with probability one.

When $p$ increases to $100$, PS fails to achieve convergence in solving score equation of $\phi$. Thus, no numerical results are presented for PS.  LASSO obtains low coverage probabilities, because of large model uncertainty. However, BSPS still works well and obtains similar performance with TPS. OBSPS outperforms by far all other methods.

Table \ref{tbl:sbps_res2_model1}  presents the numerical results for $\mathcal{M}_2$, where the outcome model is quadratic but our working model is still linear.  The same conclusion from Table \ref{tbl:sbps} can be made for the results of Table \ref{tbl:sbps_res2_model1}. OBSPS can only correctly identify $x_{i4}$ without $x_{i3}$, since $x_{i3}$ is not correlated with $y$, even though the true outcome model has $x_{i3}^2$. Overall, OBSPS is the most efficient method. The model uncertainty of LASSO keeps increasing as $p$ increase, which leads to low coverage probabilities and biased estimation for standard errors.  BSPS achieves comparable results with TPS.


\subsection{Simulation study II}
We  also apply the Bayesian sparse propensity score method to the 2006 Korean Labor and Income Panel Survey (KLIPS) data. A breif description of the panel survey can be found at http://www.kli.re.kr/klips/en/about/introduce.jsp. In KLIPS data,  there are 2,506 regular wage earners.  The study variable $y$ is the monthly income in 2006.  The auxiliary variables $( x)$ include the average monthly income in previous year and demographic variables.  We grouped age into three levels: $\text{age}<35,35\leq\text{age}<51,\text{age}\geq 51$.

 In this simulation study, we use the KLIPS data as a finite population. The realized sample is then obtained from the population by Simple Random Sampling (SRS) with sample size $n=200$ independently. Since the KLIPS data are fully observed, we artificially create a nonresponse scheme by applying the missing mechanism $\mathcal R$ in (\ref{R}). Note the two major differences here compared with the first simulation study. One is the mixed data types of the auxiliary variables. Another is the unknown outcome regression model.   The simulation process is described in the  following:
\begin{itemize}
	\item [Step 1:] Obtain 200 samples from the KLIPS data by SRS.
	\item [Step 2:] Apply the response mechanism $\mathcal R$ to the sample, so that the auxiliary variables are fully observed and the study variable $y$ is subject to missingness.
	\item [Step 3:] Apply the PS, LASSO, BSPS and OBSPS methods in simulation study I to the incomplete sample.
	\item [Step 4: ] Repeat Step 1--3 for $B=2,000$ times.
\end{itemize}

The true response function $\mathcal R$ is
\begin{eqnarray}
Pr(\delta_i=1\mid  x_i, y_i)=\frac{\exp(\phi_0+\phi_1x_{i9})}{1+\exp(\phi_0+\phi_1x_{i9})},\label{R}
\end{eqnarray}
where $(\phi_0,\phi_1)=(3,-1)$, $x_{i9}$ is average monthly income in the previous year, and the response rate is approximately 70\%. Suppose we are interested in the average monthly income $\theta=E(y)$. 
To fit the response model, we assume the response mechanism is
\begin{eqnarray}
Pr(\delta_i=1\mid  x_i, y_i)=\frac{\exp(x_i^T\phi)}{1+\exp(x_i^T\phi)}=:\pi(\phi;x_i), \nonumber
\end{eqnarray}
which is known up to the parameter $\phi$. Thus, the joint estimating equations are
\begin{eqnarray}
U_n(\phi, \theta)=\left\lbrace
\begin{array}{l}
n^{-1}\sum_{i=1}^{n}\left\lbrace \delta_i-\pi(\phi;x_i) \right\rbrace x_i\\
n^{-1}\sum_{i=1}^{n}\frac{\delta_i}{\pi(\phi;x_i)}(y_i-\theta).
\end{array}\right.\label{eq:joint_real_data}
\end{eqnarray}
The analysis result is summarized in Table \ref{KLIPS}.

\begin{table}[ht]
	\centering
	\caption{Simulation result for the 2006 Korean Labor and Income Panel Survey. ``MSE" is the mean squared error. ``Rbias" represents the relative bias of the variance estimator.}\label{KLIPS}
	\begin{tabular}{lrr}
		\hline
		Method & MSE & Rbias$\times 100$ \\ 
		\hline
		PS & 320.8 & -78.3 \\ 
		LASSO & 321.8 & -38.2 \\ 
		BSPS & 323.0 & -3.6 \\ 
		OBSPS & 311.2 & -4.4 \\ 
		\hline
	\end{tabular}
\end{table}

From Table \ref{KLIPS}, the mean square errors of all four methods are similar. However, the proposed optimal Bayesian sparse propensity score method (OBSPS) is most efficient, because OBSPS incorporates the relevant auxiliary variables in propensity score estimation. Due to large dimensions of auxiliary variables, the traditional propensity score (PS) estimation including all variables fails to provide a consistent variance estimator, as explained in Theorem \ref{high_PS}.  The propensity score model using LASSO also highly underestimates the variance due to large model uncertainty. In summary, the proposed BSPS and OBSPS provide consistent variance estimators uniformly regardless of the dimension of covariates.


\section{Discussion}\label{sec::discussion}
This paper presents a Bayesian approach to PS estimation using the Spike-and-Slab prior for the response propensity model. Through the proposed BSPS method, model selection consistency holds and the uncertainty in model selection is fully captured by the Bayesian framework. The efficiency of the PS estimation can be further improved by incorporating relevant auxiliary variables, the so-called optimal Bayesian sparse propensity score (OBSPS) method. The simulation study in Section \ref{sec::simulation} shows that
the Bayesian approach provides valid frequentist coverage probabilities in finite samples. Since the PS estimation is widely used in causal inference \citep{morgan2014counterfactuals,hudgens2008toward}, applying the proposed methods to the sparse Bayesian causal inference can be developed similarly. Also, our proposed method is developed under the assumption of MAR. Extension of our proposed method to nonignorable nonresponse is a topic for future research.

\clearpage
\appendix
{\LARGE \bf Appendices}

\vspace{0.01\textheight}

In these appendices, we present the technical derivations and proofs for all stated theorems in this paper.

\section{Computational details} \label{App::A}

 To generate $\phi^{(t+1)}$ from (\ref{post_phi1}), the computation using the Metropolis-Hastings algorithm \citep{chib1995understanding} can be quite heavy. Thus, instead of using the likelihood function of $\phi$ directly, we propose to use the Laplace approximation method. To discuss the approximation of (\ref{post_phi1}), let $\hat \phi^{(t+1)}$ be the maximizer of $L_1(\phi\mid \mbox{data})p(\phi\mid z^{(t+1)})$.
From the Spike-and-Slab prior in (\ref{ss:prior}), $p(\phi\mid z^{(t+1)})$ is a Gaussian distribution with mean 0 and variance  $V_{z^{(t+1)}}=\text{Diag}\left(\nu_{z_1^{(t+1)}},\nu_{z_2^{(t+1)}},\ldots,\nu_{z_p^{(t+1)}}\right)$, where $\nu_{z_j^{(t+1)}}=\nu_1z^{(t+1)}_j+\nu_0(1-z^{(t+1)}_j)$. Thus, maximizing $L_1(\phi\mid \mbox{data})p(\phi\mid z^{(t+1)})$ is equivalent to solving
	\begin{eqnarray}
	S_n(\phi)-V^{-1}_{z^{(t+1)}}\phi=0.\label{laplace_eq}
	\end{eqnarray}
Denote \begin{eqnarray}
\hat V_{\phi}=\left( nI_{\phi}+V^{-1}_{z^{(t+1)}}\right)^{-1},\label{V_phi}
\end{eqnarray}
where $I_{\phi}$ is the negative fisher information matrix of $\phi$ defined as
\begin{eqnarray}
I_{\phi}=E\left\lbrace \frac{\partial^2\log f(\delta_i\mid x_i, \phi)}{\partial \phi \partial \phi^T}\right\rbrace.\nonumber
\end{eqnarray}
Note that $V_{\phi}$ is always positive definite. Using second oder Taylor expansion, the Laplace approximation is
\begin{eqnarray}
&L_1(\phi\mid \mbox{data})p(\phi\mid z^{(t+1)})\cong &L_1(\hat \phi^{(t+1)}\mid \mbox{data})p(\hat \phi^{(t+1)}\mid z^{(t+1)})\nonumber\\
&&\times\exp\left\lbrace -\frac{1}{2}(\phi-\hat \phi^{(t+1)})^T \hat V^{-1}_{\hat \phi^{(t+1)}}(\phi-\hat \phi^{(t+1)})\right\rbrace .\nonumber
\end{eqnarray}
Therefore,  generating $\phi^{(t+1)}$ from (\ref{post_phi1}) is approximately equivalent to generating $\phi^{(t+1)}$ from $N(\hat \phi^{(t+1)}, \hat V_{\hat \phi^{(t+1)}})$, where
$\hat V_{\hat \phi^{(t+1)}}$ is a consistent estimator with plugged in $\hat I_{\hat \phi^{(t+1)}}$.

For \emph{Step} 2b, note that,  under some regularity conditions, we can establish that
	\begin{eqnarray}
	n^{-1/2} \left.\left\lbrace \begin{matrix}S_n \left( \phi_{z^{(t+1)}}\right)\\
	U_{PS}\left(\theta, \phi_{z^{(t+1)}} \right)
	\end{matrix}\right\rbrace   \right|\phi_{{z^{(t+1)}}},\theta  \stackrel{\mathcal{L}}{ \longrightarrow}  N\left( \bm 0, \Sigma^{(t+1)} \right). \label{asym2}
	\end{eqnarray}
	Correspondingly, $\Sigma^{(t+1)}=\Sigma^{(t+1)}( \phi_{z^{(t+1)}}, \theta)$ can be decomposed as
	\begin{eqnarray}
	\Sigma^{(t+1)}=\left( \begin{matrix}
	\Sigma^{(t+1)}_{11} & \Sigma^{(t+1)}_{12}\\
	\Sigma^{(t+1)}_{21} & \Sigma^{(t+1)}_{22}
	\end{matrix}\right).\nonumber
	\end{eqnarray}
	Thus, the asymptotic distribution in (\ref{asym2}) implies that $\sqrt{n} U_{PS}(\theta,\phi_{z^{(t+1)}})\mid   S_n(\phi_{z^{(t+1)}}),  \theta, \phi_{z^{(t+1)}}^{(t+1)}$ goes to a normal distribution  $$N\left\lbrace \Sigma^{(t+1)}_{21}\left(  \Sigma^{(t+1)}_{11} \right)^{-1} S_n(\phi_{z^{(t+1)}}^{(t+1)}), \Sigma^{(t+1)}_{22\cdot 1}\right\rbrace ,$$ where $\Sigma^{(t+1)}=\Sigma^{(t+1)}( \phi_{z^{(t+1)}}^{(t+1)}, \theta)$ and $\Sigma^{(t+1)}_{22\cdot 1}=\Sigma^{(t+1)}_{22}-\Sigma^{(t+1)}_{21}\left( \Sigma^{(t+1)}_{11}\right)^{-1} \Sigma^{(t+1)}_{12} $. Therefore, $g_2\left\lbrace   U_{PS}(\theta,\phi_{z^{(t+1)}})\mid   S_n(\phi_{z^{(t+1)}}),  \theta, \phi_{z^{(t+1)}}^{(t+1)}\right\rbrace$ is a normal density function.

To establish consistency, we assume the following condition to avoid unnecessary details :
\begin{itemize}
	\item[]  (\textbf{A5}) The $\hat V_{\phi}$  in (\ref{V_phi}) satisfies $\hat V_{ \phi}=V_{\phi}\left\{ 1+o_p(1)\right\} $. \label{asump:a7}
\end{itemize}

\section{Implementation of the \textit{P-step}}\label{App::B}
Given $u^{(t+1)}, \sigma_e^{2,(t)}$, generate $\beta^{(t+1)}$ from a multivariate Gaussian distribution with mean $\mu^*$ and variance $V^*$, where
\begin{eqnarray}
V^*=\left( V^{-1}_{u^{(t+1)}}+\frac{\sum_{i=1}^n\delta_i x_ix_i^\T}{\sigma_e^{2,(t)}}\right)^{-1},\nonumber\\
\mu^*=\left( V^{-1}_{u^{(t+1)}}+\frac{\sum_{i=1}^n\delta_i x_ix_i^\T}{\sigma_e^{2,(t)}}\right)^{-1}\frac{\sum_{i=1}\delta_i x_i y_i}{\sigma_e^{2,(t)}},\label{Pstep1}
\end{eqnarray}
and $V_{u^{(t+1)}}=\mathrm{Diag}\left\lbrace \gamma_{u_1^{(t+1)}}, \cdots, \gamma_{u_p^{(t+1)}}\right\rbrace $. Then, given $\beta^{(t+1)}$, $\sigma_e^{2, (t+1)}$ is generated from a inverse gamma distribution with parameters $(c_1^*, c_2^*)$, where
\begin{eqnarray}
c_1^*=c_1+\frac{r}{2},\nonumber\\
c_2^*=c_2+\frac{1}{2}\sum_{i=1}^{n}\delta_i (y_i-x_i^\T\beta^{(t+1)})^2.\label{Pstep2}
\end{eqnarray}

\section{ Proof of Theorem \ref{high_PS}}\label{proof_theorem1}
 Without loss of generality, we assume
\begin{eqnarray}
E\left( XX^\T\right)=\left( \begin{matrix}
E\left( X_1X_1^\T\right) & 0 & 0\\
0 & E(X_2X_2^\T) & 0\\
0 & 0 &E(X_3X_3^\T)
\end{matrix}\right) \label{A1}
\end{eqnarray}
to simplify the proof.

Assume $\hat \eta=(\hat \phi, \hat \theta)$ is the solution of
\begin{eqnarray}
U_n(\eta)=\left\lbrace \begin{matrix}
n^{-1}\sum_{i=1}^{n} S(\phi; x_i, \delta)\\
n^{-1}\sum_{i=1}^n \delta_i \pi^{-1}(\phi; x_i) U(\theta; x_i, y_i)\\
\end{matrix}\right\rbrace ,\nonumber
\end{eqnarray}
where $\pi(\phi; x_i)=G(x_i^\T\phi)$.
Thus, we can derive the score function of $\phi$ as
\begin{eqnarray}
S(\phi; x_i, \delta_i)=\left\lbrace \frac{\delta_i}{G(x_i^\T\phi)}-\frac{1-\delta_i}{1-G(x_i^\T\phi)} \right\rbrace G'(x_i^\T \phi) x_i,\nonumber
\end{eqnarray}
where $G'(\cdot)$ is the first order derivative of $G(\cdot)$.

We first consider that $p_3=O(1)$.
Applying the Taylor expansion to the joint estimating equations, we have
\begin{eqnarray}
U_n(\hat \eta)=U_n(\eta_0)+E\left\lbrace \left.\frac{\partial U_n(\eta)}{\partial \eta^\T}\right|_{\eta=\eta_0}\right\rbrace (\hat \eta-\eta_0)+O_p(\|\hat \eta-\eta_0\|^2).\nonumber
\end{eqnarray}
If $p_3=O(1)$, we can ignore the smaller term and obtain
\begin{eqnarray}
\hat \eta-\eta_0= -\left[ E\left\lbrace \left.\frac{\partial U_n(\eta)}{\partial \eta^\T}\right|_{\eta=\eta_0}\right\rbrace\right]^{-1} U_n(\eta_0).\nonumber
\end{eqnarray}
Then, the variance is asymptotically equal to
\begin{eqnarray}
\mathrm{var}\left(\hat \eta-\eta_0 \right) =\left[ E\left\lbrace \left.\frac{\partial U_n(\eta)}{\partial \eta^\T}\right|_{\eta=\eta_0}\right\rbrace\right]^{-1}\mathrm{var}\left\lbrace U_n(\eta_0) \right\rbrace \left[ E\left\lbrace \left.\frac{\partial U_n(\eta)}{\partial \eta^\T}\right|_{\eta=\eta_0}\right\rbrace\right]^{-1,\T}.\label{var}
\end{eqnarray}
Now, let us compute the variance in (\ref{var}). First, we can show that
\begin{eqnarray}
E\left\lbrace \left.\frac{\partial U_n(\eta)}{\partial \eta^\T}\right|_{\eta=\eta_0}\right\rbrace=\left( \begin{matrix}
A & 0\\
C & D
\end{matrix}\right),\nonumber
\end{eqnarray}
where
\begin{eqnarray}
A=-E\left( \left[ G^{-1}(X^\T\phi_0)+\left\lbrace 1-G(X^\T\phi_0)\right\rbrace^{-1} \right] G'(X^\T\phi_0)G'(X^\T\phi_0) XX^\T\right) ,\nonumber\\
C=-E\left\lbrace G^{-1}(X^\T\phi_0)G^{'}(X^\T\phi_0)U(\theta_0; X,Y)X^\T\right\rbrace ,\nonumber\\
D=E\left\lbrace \frac{\partial U(\theta_0; X, Y)}{\partial \theta}\right\rbrace.\nonumber
\end{eqnarray}
Under the true model assumption, we have $G(X^\T\phi_0)=G(X_1^\T\phi_{0,1})=G_0(X_1)$, where $\phi_0=(\phi_{0,1}, 0, 0)$. Moreover, we can decompose $A$ as
\begin{eqnarray}
A=\left( \begin{matrix}
A_1 & 0 & 0\\
0 & A_2& 0\\
0 & 0 & A_3
\end{matrix}\right) ,
\end{eqnarray}
where
\begin{eqnarray}
A_{1}=-E\left( \left[ G_0^{-1}(X_1)+\left\lbrace 1-G_0(X_1)\right\rbrace^{-1} \right] G'_0(X_1)G'_0(X_1) X_1X_1^\T\right) ,\nonumber\\
A_2=-E\left( \left[ G_0^{-1}(X_1)+\left\lbrace 1-G_0(X_1)\right\rbrace^{-1} \right] G'_0(X_1)G'_0(X_1)\right)  E\left( X_2X_2^\T\right) ,\nonumber\\
A_3=-E\left( \left[ G_0^{-1}(X_1)+\left\lbrace 1-G_0(X_1)\right\rbrace^{-1} \right] G'_0(X_1)G'_0(X_1)\right) E \left( X_3X_3^\T\right).\nonumber
\end{eqnarray}
Similarly, we can show that
\begin{eqnarray}
&C&=\left( -E\left\lbrace G_0^{-1}(X_1)G^{'}_0(X_1)U(\theta_0; X,Y)X_1^\T\right\rbrace,-E\left\lbrace G_0^{-1}(X_1)G^{'}_0(X_1)U(\theta_0; X,Y) X_2^\T\right\rbrace,0\right) \nonumber\\
&&=:(C1,C_2,0).\nonumber
\end{eqnarray}

Then, we derive the variance of $U_n(\eta_0)$ as
\begin{eqnarray}
&\mathrm{var}\left\lbrace U_n(\eta_0)\right\rbrace &=E\left\lbrace U_n(\eta_0)U^T_n(\eta_0) \right\rbrace \nonumber\\
&&= n^{-1}\left( \begin{matrix}
-A & -C^\T\\
- C & E\left\lbrace U^2(\theta_0; X,Y)\right\rbrace
\end{matrix}\right)
\end{eqnarray}

To compute the variance of $\hat \eta$, we apply block matrix inverse formula to
\begin{eqnarray}
E\left\lbrace \left.\frac{\partial U_n(\eta)}{\partial \eta^\T}\right|_{\eta=\eta_0}\right\rbrace.\nonumber
\end{eqnarray}
That is
\begin{eqnarray}
\left( \begin{matrix}
A & 0\\
C & D
\end{matrix}\right)^{-1}=\left( \begin{matrix}
A^{-1} & 0\\
-D^{-1} CA^{-1} & D^{-1}
\end{matrix}\right).\nonumber
\end{eqnarray}
Finally, we can obtain that
\begin{eqnarray}
&\mathrm{var}\left(\hat \eta-\eta_0 \right) &=n^{-1}\left( \begin{matrix}
A^{-1} & 0\\
-D^{-1} CA^{-1} & D^{-1}
\end{matrix}\right)\left( \begin{matrix}
-A & -C^\T\\
-C & E\left\lbrace U^2(\theta_0; X,Y)\right\rbrace
\end{matrix}\right) \left( \begin{matrix}
A^{-1} & -A^{-1}C^\T D^{-1}\\
0 & D^{-1}
\end{matrix}\right)\nonumber\\
&&=n^{-1}\left(\begin{matrix}
-A^{-1} & 0\\
0 & D^{-1}E\left\lbrace U^2(\theta_0; X,Y)\right\rbrace D^{-1}+D^{-1} CA^{-1} C^{-1} D^{-1}
\end{matrix} \right)=:\Sigma .\nonumber
\end{eqnarray}

Therefore, we have
\begin{eqnarray}
&\mathrm{var}(\hat \theta_{PS})&=n^{-1} E\left\lbrace \frac{\partial U(\theta_0; X, Y)}{\partial \theta}\right\rbrace E\left\lbrace U^2(\theta_0; X,Y)\right\rbrace E\left\lbrace \frac{\partial U(\theta_0; X, Y)}{\partial \theta}\right\rbrace \nonumber\\
&&+E\left\lbrace \frac{\partial U(\theta_0; X, Y)}{\partial \theta}\right\rbrace\left(C_1A_1^{-1}C_1^\T +C_2A_2^{-1}C_2^\T \right)  E\left\lbrace \frac{\partial U(\theta_0; X, Y)}{\partial \theta}\right\rbrace.\nonumber
\end{eqnarray}

Since $A_1, A_2$ is negative definite,
\begin{eqnarray}
E\left\lbrace \frac{\partial U(\theta_0; X, Y)}{\partial \theta}\right\rbrace\left(C_1A_1^{-1}C_1^\T \right)  E\left\lbrace \frac{\partial U(\theta_0; X, Y)}{\partial \theta}\right\rbrace\leq 0\nonumber\\
E\left\lbrace \frac{\partial U(\theta_0; X, Y)}{\partial \theta}\right\rbrace\left(C_2A_2^{-1}C_2^\T \right)  E\left\lbrace \frac{\partial U(\theta_0; X, Y)}{\partial \theta}\right\rbrace\leq 0\nonumber.\nonumber
\end{eqnarray}

Therefore, the PS estimator using the true response probability is less efficient than the PS estimator using the true response model with estimated response probability. Moreover, the PS estimator using estimated response probability in the true response model is less efficient than the PS estimator using estimated response probability including $X_2$. This completes the proof for the last part of Theorem \ref{high_PS}.

Then, we consider $p_3=p_3(n)$ case. Under this case, $O_p(\|\hat \eta-\eta_0\|^2)$ is not negligible. We expand $U_n(\hat \eta)$ to the second order term in Taylor expansion. That is
\begin{eqnarray}
&U_n(\hat \eta)&=U_n(\eta_0)+E\left\lbrace \left.\frac{\partial U_n(\eta)}{\partial \eta^\T}\right|_{\eta=\eta_0}\right\rbrace (\hat \eta-\eta_0)+R +O_p(\|\hat \eta-\eta_0\|^3),\nonumber
\end{eqnarray}
where $R=(R_1, \cdots, R_{p+1})$, $R_j=\frac{1}{2}\sum_{j=1}^p (\hat \eta-\eta_0)^\T H_j(\hat \eta-\eta_0)$ and $H_j=E\left\lbrace \left. \partial^2 U_n(\eta)/\partial \eta^\T \partial \eta_j\right|_{\eta=\eta_0} \right\rbrace $.
Thus,
\begin{eqnarray}
E\left( \hat \eta -\eta_0\right) \cong -\left[ E\left\lbrace \left.\frac{\partial U_n(\eta)}{\partial \eta^\T}\right|_{\eta=\eta_0}\right\rbrace\right]^{-1} E(R).
\end{eqnarray}

Assume $$ E\left\lbrace \left.\frac{\partial U_n(\eta)}{\partial \eta^\T}\right|_{\eta=\eta_0}\right\rbrace=O(1).$$
Moreover, assume $H_j=O(1)$. Then we have
\begin{eqnarray}
&E(R_j)&=O\left[ E\left\lbrace (\hat \eta-\eta_0)^\T H_j(\hat \eta-\eta_0)\right\rbrace \right] \nonumber\\
&&=O\left(  \mathrm{Trace}\left[ E\left\lbrace H_j(\hat \eta-\eta_0)(\hat \eta-\eta_0)^\T \right\rbrace \right] \right) \nonumber\\
&&=O\left(  \mathrm{Trace}\left[ E\left\lbrace (\hat \eta-\eta_0)(\hat \eta-\eta_0)^\T \right\rbrace \right] \right)\nonumber\\
&&=O\left[  n^{-1}\left\lbrace \mathrm{Trace}(A_1^{-1}) +\mathrm{Trace}(A_2^{-1}) +\mathrm{Trace}(A_3^{-1}) \right\rbrace \right]\nonumber\\
&&=O\left(\frac{p_3}{n} \right) ,\nonumber
\end{eqnarray}
which leads to $\mu=E\left( \hat \theta_{PS}-\theta_0\right)=O(p_3/n)=O(p/n)$.

Similarly,
\begin{eqnarray}
&\mathrm{var}\left( R_j\right) &=O\left[ \mathrm{var}\left\lbrace (\hat \eta-\eta_0)^\T H_j(\hat \eta-\eta_0)\right\rbrace \right] \nonumber\\
&&=O\left\lbrace \mathrm{Trace}\left( H_j\Sigma H_j\Sigma\right) \right\rbrace +O\left( \mu^TH_j\Sigma H_j\mu\right) \nonumber\\
&&=O\left\lbrace \mathrm{Trace}(\Sigma\Sigma)\right\rbrace +O(\mu^T\Sigma\mu) \nonumber\\
&&=O\left\lbrace n^{-2}\mathrm{Trace}(A^{-1})\right\rbrace +O\left( \frac{p_3^2}{n^3}\right) \nonumber\\
&&=O\left( \frac{p_3}{n^2}\right)+O\left( \frac{p_3^2}{n^3}\right) .\label{cc}
\end{eqnarray}
Let $\hat \eta_j=\hat \theta_{PS}$ and we have $$\mathrm{var}\left( \hat \theta_{PS}-\theta_0\right) =O(1/n)+O\left\lbrace p_3n^{-2}(1+p_3/n)\right\rbrace .$$
We complete the proof for Theorem \ref{high_PS}.

\section{ Proof of Theorem \ref{thm:model_consistnecy}}\label{proof_theorem2}

Let $V=n^{-1}\mathrm{I}^{-1}_{\phi_0}$ in (\ref{V_phi}).
Our proof can be summarized as follows:
First, we show that
\begin{eqnarray}\label{claim:B1}
\tilde{p}(z_o|\mbox{data})\overset{p}{\to} 1,
\end{eqnarray}
as $n\to \infty$, where $$\tilde{p}(z_o|\mbox{data})=\frac{\int  \psi(\hat{\phi}|\phi,V) p(\phi|z_o) p(z_o)   d\phi}{ \int \int \psi(\hat{\phi}|\phi,V)  p(\phi|z) p(z) d\phi dz },$$
 $\psi(\cdot\mid \phi, V)$ is the normal density function with mean $\phi$ and variance $V$, and $\hat \phi$ is the maximizer of $L_1(\phi\mid \mbox{data})$.

Second, we show that
\begin{eqnarray}\label{claim:B2}
\left|\tilde{p}(z_o|\mbox{data})-p_g(z_o|\mbox{data})\right|\overset{p}{\to} 0,
\end{eqnarray}
as $n\to \infty$.
Note that
$$\left|\tilde{p}(z_o|\mbox{data})-p_g(z_o|\mbox{data})\right|\geq \left|\left|\tilde{p}(z_o|\mbox{data})-1\right|-\left|p_g(z_o|\mbox{data})-1\right|\right|.$$
Finally, by (\ref{claim:B1}) and (\ref{claim:B2}), we have that
\begin{eqnarray*}
	p_g(z_o|\mbox{data})\overset{p}{\to} 1,
\end{eqnarray*}
as $n\to \infty$.

\newpage

\noindent {\bf  Proof of Claim (\ref{claim:B1})}

Under (\textbf{A4}), since $\pi(z)\propto 1$, $\tilde{p}(z_o|\mbox{data})$ reduces to
\begin{eqnarray*}
	\tilde{p}(z_o|\mbox{data})&=&\frac{\int  \psi(\hat{\phi}|\phi,V) p(\phi|z_o)  d\phi}{ \sum_{z\in\{0,1\}^{p}} \int \psi(\hat{\phi}|\phi,V)  p(\phi|z) d\phi }\\
	&:=&\frac{f(\hat{\phi}|z_o)} {\sum_{z\in\{0,1\}^{p}} f(\hat{\phi}|z) }\\
	&=&\frac{1} {1+\sum_{z\neq z_o} \frac{f(\hat{\phi}|z)}{f(\hat{\phi}|z_o)} },
\end{eqnarray*}
where $f(\hat{\phi}|z)=\int  \psi(\hat{\phi}|\phi,V) p(\phi|z)  d\phi $. Our proof can be done by showing that
\begin{eqnarray}
\sum_{z\neq z_o}\frac{f(\hat{\phi}|z)}{f(\hat{\phi}|z_o)}\overset{p}{\to} 0,\end{eqnarray}
as $n\to \infty$. Since $\Sigma=\mathrm{I}_{\phi_0}^{-1}$ is symmetric and positive definite, by spectral decomposition, $\Sigma$ can be factorized as $\Sigma=Q\Lambda Q^{-1}$, where $\Lambda$ is the diagonal matrix whose diagonal elements are the eigenvalues of $\Sigma$ and each column of $Q$ is the eigenvector of $\Sigma$. Since $V=n^{-1}\Sigma$, we have $V=Q(n^{-1}\Lambda)Q^{-1}$. Let $\lambda_{n,\min}=n^{-1}\lambda_{\min}$ and $\lambda_{n,\max}=n^{-1}\lambda_{\max}$, where $\lambda_{\min}$ and $\lambda_{\max}$ indicate the smallest and the largest diagonal elements of $\Lambda$, respectively. Note that $\lambda_{n,\min}^{-1}I -V^{-1}$ and $V^{-1}-\lambda_{n,\max}I$ are positive semidefinite due to the fact that
\begin{eqnarray}
\lambda_{n,\min}^{-1}I -V^{-1}=Q\left(\lambda_{n,\min}^{-1}I-n{\Lambda}^{-1} \right) Q^{-1} \nonumber,\\
V^{-1}-\lambda_{n,\max}^{-1}I=Q\left(n {\Lambda}^{-1}-\lambda_{n,\max}^{-1}  I\right) Q^{-1} \nonumber.
\end{eqnarray}
This implies that
\begin{eqnarray}\label{eq.B7}
\lambda_{n,\max}^{-1} w^\T w\leq  w^\T V^{-1} w \leq \lambda_{n,\min}^{-1} w^\T w,
\end{eqnarray}
for any $w$. Recall that
\begin{eqnarray}
\psi(\hat{\phi}|\phi,V)=c\exp\left\lbrace -\frac{1}{2} \left(\hat \phi-\phi \right)^\T  V^{-1}\left(\hat \phi-\phi \right)   \right\rbrace ,\nonumber
\end{eqnarray}
where $c$ denotes the normalizing constant. From (\ref{eq.B7}), we have
\begin{eqnarray}
\psi(\hat{\phi}|\phi,V)\geq c\exp\left\lbrace -\sum_{j=1}^{p} \frac{1}{2\lambda_{n,\min}}\left(\hat \phi_j-\phi_j \right)^2 \right\rbrace , \label{eq.B8}\\
\psi(\hat{\phi}|\phi,V)\leq c\exp\left\lbrace -\sum_{j=1}^{p} \frac{1}{2\lambda_{n,\max}}\left(\hat \phi_j-\phi_j \right)^2 \right\rbrace.\label{eq.B9}
\end{eqnarray}
Using (\ref{eq.B8}), we construct a lower bound of $f(\hat{\phi}|z)=\int  \psi(\hat{\phi}|\phi,V) p(\phi|z)  d\phi  $ as
\begin{eqnarray*}
	f(\hat{\phi}|z)&\geq& c \prod_{j=1}^{p} \left(2\pi \nu_{z_j} \right)^{-1/2}\int \exp\left\lbrace -\frac{1}{2\lambda_{n,\min}}\left(\hat \phi_j-\phi_j \right)^2-\frac{1}{2\nu_{z_j}}\phi_j^2 \right\rbrace d\phi_j  \\
	&=& c_2\prod_{j=1}^{p}\left(\frac{\lambda_{n,\min}}{\lambda_{n,\min}+\nu_{z_j}} \right)^{1/2}\exp\left\lbrace -\frac{\hat \phi_j^2}{2\left( \lambda_{n,\min}+\nu_{z_j}\right) } \right\rbrace ~\equiv L_f(z). \nonumber
\end{eqnarray*}
Similarly, using (\ref{eq.B9}), we construct an upper bound of $f(\hat{\phi}|z)$ as
\begin{eqnarray*}
	f(\hat{\phi}|z)&\leq& c_3\prod_{j=1}^{p}\left(\frac{\lambda_{n,\max}}{\lambda_{n,\max}+\nu_{z_j}} \right)^{1/2}\exp\left\lbrace -\frac{\hat \phi_j^2}{2\left( \lambda_{n,\max}+\nu_{z_j}\right) } \right\rbrace ~\equiv U_f(z).\nonumber
\end{eqnarray*}
Hence, we have
\begin{eqnarray}
\frac{f(\hat{\phi}|z)}{f(\hat{\phi}|z_o)}\leq \frac{U_f(z)}{L_f(z_o)}.
\end{eqnarray}
We now claim $ \frac{U_f(z)}{L_f(z_o)}\overset{p}{\to} 0$ as $n\to 0$ for any $z\neq z_o$.
Define
$$H_n(z_j,z_{o,j})= \left\{\frac{\lambda_{n,\max}(\lambda_{n,\min}+\nu_{z_{o,j}})}{\lambda_{n,\min}(\lambda_{n,\max}+\nu_{z_j})} \right\}^{1/2}\exp\left\lbrace -\frac{\hat \phi_j^2}{2( \lambda_{n,\max}+\nu_{z_j}) }+\frac{\hat \phi_j^2}{2( \lambda_{n,\min}+\nu_{z_{o,j}}) }\right\rbrace.$$ Suppose $z_{o,j}=0$. Then we have that $\hat{\phi}_j^2=O_p(n^{-1})$ from Theorem \ref{high_PS}. Recall that from (\textbf{A4}), $\nu_0=o(n^{-1})$. If $z_j=0$, then
\begin{eqnarray*}
	&&H_n(0,0)= \left\{\frac{\lambda_{n,\max}(\lambda_{n,\min}+\nu_{0})}{\lambda_{n,\min}(\lambda_{n,\max}+\nu_{0})} \right\}^{1/2}\exp\left\lbrace -\frac{\hat \phi_j^2}{2( \lambda_{n,\max}+\nu_{0}) }+\frac{\hat \phi_j^2}{2( \lambda_{n,\min}+\nu_{0}) }\right\rbrace\\
	&&= \left\{\frac{O(n^{-2})+o(n^{-2}) }{O(n^{-2})+o(n^{-2})} \right\}^{1/2} \exp\left\lbrace -\frac{O_p(n^{-1}) }{O(n^{-1}) +o(n^{-1})}+\frac{O_p(n^{-1})}{ O(n^{-1})+o(n^{-1})}\right\rbrace.
\end{eqnarray*}
This implies that $H_n(0,0)=1$ in probability. From (\textbf{A4}), we have $\nu_1=O(n)$. If $z_j=1$, then
\begin{eqnarray*}
	&&H_n(1,0)= \left\{\frac{\lambda_{n,\max}(\lambda_{n,\min}+\nu_{0})}{\lambda_{n,\min}(\lambda_{n,\max}+\nu_{1})} \right\}^{1/2}\exp\left\lbrace -\frac{\hat \phi_j^2}{2( \lambda_{n,\max}+\nu_{1}) }+\frac{\hat \phi_j^2}{2( \lambda_{n,\min}+\nu_{0}) }\right\rbrace\\
	&&= \left\{\frac{O(n^{-2})+o(n^{-2}) }{O(n^{-2})+O(1)} \right\}^{1/2} \exp\left\lbrace -\frac{O_p(n^{-1}) }{2\{O(n^{-1}) +O(n)\} }+\frac{O_p(n^{-1})}{2\{ O(n^{-1})+o(n^{-1})\} }\right\rbrace.
\end{eqnarray*}
This implies that $H_n(1,0)=O_p(n^{-1})$. Suppose $z_{o,j}=1$. Then we have $\hat{\phi}_j =O_p(1)$. If $z_j=0$, then
\begin{eqnarray*}
	&&H_n(0,1)= \left\{\frac{\lambda_{n,\max}(\lambda_{n,\min}+\nu_{1})}{\lambda_{n,\min}(\lambda_{n,\max}+\nu_{0})} \right\}^{1/2}\exp\left\lbrace -\frac{\hat \phi_j^2}{2( \lambda_{n,\max}+\nu_{0}) }+\frac{\hat \phi_j^2}{2( \lambda_{n,\min}+\nu_{1}) }\right\rbrace\\
	&&= \left\{\frac{O(n^{-2})+O(1) }{O(n^{-2})+o(n^{-2})} \right\}^{1/2} \exp\left\lbrace -\frac{O_p(1)}{2\{O(n^{-1}) +o(n^{-1})\} }+\frac{O_p(1) }{2\{ O(n^{-1})+O(n)\} }\right\rbrace\\
	&&= O(n)  \exp\left\lbrace - O_p(n) \right\rbrace.
\end{eqnarray*}
This implies that $H_n(0,1)=O_p\{\exp(-n)\}$. When $z_j=1$, we have
\begin{eqnarray*}
	&&H_n(1,1)= \left\{\frac{\lambda_{n,\max}(\lambda_{n,\min}+\nu_{1})}{\lambda_{n,\min}(\lambda_{n,\max}+\nu_{1})} \right\}^{1/2}\exp\left\lbrace -\frac{\hat \phi_j^2}{2( \lambda_{n,\max}+\nu_{1}) }+\frac{\hat \phi_j^2}{2( \lambda_{n,\min}+\nu_{1}) }\right\rbrace\\
	&&= \left\{\frac{O(n^{-2})+O(1) }{O(n^{-2})+O(1)} \right\}^{1/2} \exp\left\lbrace -\frac{O_p(1) }{2\{O(n^{-1}) +O(n)\} }+\frac{O_p(1) }{2\{ O(n^{-1})+O(n)\} }\right\rbrace.
\end{eqnarray*}
This implies that $H_n(1,1)=O_p(1)$. Note that
\begin{eqnarray*}\label{UB}
	\frac{U_f(z)}{L_f(z_o)}\propto \prod_{j=1}^p H_n(z_j,z_{o,j}).
\end{eqnarray*}
If $z\neq z_o$, then $ \prod_{j=1}^p H_n(z_j,z_{o,j})$ must include at least one of $H_n(1,0)$ or $H_n(0,1)$.

Note that
\begin{eqnarray}
&\sum_{z\neq z_o}\frac{f(\hat{\phi}|z)}{f(\hat{\phi}|z_o)}&\leq c_4 \sum_{z\neq z_o} \frac{U_f(z)}{L_f(z_o)}\nonumber\\
&&\leq c_4\sum_{j_1\leq p_1, j_2\leq p-p_1,j_3\leq p_1, j_4\leq n-p_1, j_1+j_2+j_3+j_4=p, j_2+j_3>0} H^{j_1}_n(1,1)H^{j_2}_n(1,0)H^{j_3}_n(0,1)H^{j_4}_n(0,0)\nonumber. \\
&&\leq c_4 H^{p_1}_n(1,1) \sum_{ j_2\leq p-p_1,j_3\leq p_1, j_4\leq n-p_1, j_2+j_3+j_4=p, j_2+j_3>0}H^{j_2}_n(1,0)H^{j_3}_n(0,1)H^{j_4}_n(0,0).\nonumber
\end{eqnarray}
Since we have shown that $H_n(1,1)=H_n(0,0)=O_p(1)$, $H_n(1,0)=O_p(n^{-1})$ and $H_n(0,1)=O_p(\exp(-n))$,
we can show that $H^{p_1}_n(1,1) =O_p(1)$  and
\begin{eqnarray}
&&\sum_{ j_2\leq p-p_1,j_3\leq p_1, j_4\leq n-p_1, j_2+j_3+j_4=p, j_2+j_3>0}H^{j_2}_n(1,0)H^{j_3}_n(0,1)H^{j_4}_n(0,0)\nonumber\\
&&\leq \sum_{ j_2\leq p-p_1, j_4\leq n-p_1, j_2+j_4=p, j_2>0}H^{j_2}_n(1,0)H^{j_4}_n(0,0).\nonumber\\
&&\leq \left\lbrace H_n(0,0)+H_n(1,0)\right\rbrace^p- H^{p}_n(0,0)\nonumber\\
&&=\left\lbrace1+O_p(n^{-1}) \right\rbrace ^p-1.\nonumber
\end{eqnarray}

We know that
\begin{eqnarray}
\lim\limits_{n\infty} (1+an^{-1})^n=e^a,\nonumber
\end{eqnarray}
for any $a>0$. Thus,
\begin{eqnarray}
\left\lbrace1+O_p(n^{-1}) \right\rbrace ^p\xrightarrow{}1
\end{eqnarray}
in probability, if $p=o(n)$.
This implies that
\begin{eqnarray}
\sum_{z\neq z_o}\frac{f(\hat{\phi}|z)}{f(\hat{\phi}|z_o)}\xrightarrow{}0,\nonumber
\end{eqnarray}
in probability. This completes our proof.

\noindent {\bf  Proof of Claim (\ref{claim:B2})}

First, we show that 
\begin{eqnarray}
\psi ( \hat \phi| \phi, \hat{V})=\psi ( \hat \phi| \phi, V)\{1+o_p(1)\} , \nonumber
\end{eqnarray}
where $\hat{ V}=n^{-1} \hat \Sigma$. In (\textbf{A5}), we have
\begin{eqnarray*}
	\hat \Sigma=\Sigma\left\{1+o_p(1) \right\}.
\end{eqnarray*}
Under $\Sigma>0$, 
\begin{eqnarray*}
	|\hat \Sigma|^{-1/2}=|\Sigma|^{-1/2}\{1+o_p(1) \}.
\end{eqnarray*}
Therefore, we have
\begin{eqnarray*}
	\psi ( \hat \phi| \phi, \hat{V})= \frac{1}{(2\pi)^{\frac{p}{2}}|V|^{\frac{1}{2}}}\exp\left[-\frac{1}{2}\left(\hat\phi-\phi \right)^\T V^{-1}\left(\hat\phi-\phi \right)\{1+o_p(1) \} \right] \{1+o_p(1) \}.\label{claim:app_equal}
\end{eqnarray*}
To complete the proof, we need to show that
\begin{eqnarray}\label{arg:lim}
\exp\left[-\frac{1}{2}\left(\hat\phi-\phi \right)^\T V^{-1}\left(\hat\phi-\phi \right)o_p(1) \right]=O_p(1).
\end{eqnarray}
From (\ref{eq.B7}), we have
\begin{eqnarray*}
	\frac{n}{2\lambda_{\max}}\|\hat\phi-\phi \|^2 \leq \frac{1}{2}\left(\hat\phi-\phi \right)^\T V^{-1}\left(\hat\phi-\phi \right)\leq \frac{n}{2\lambda_{\min}}\|\hat\phi-\phi \|^2 ,
\end{eqnarray*}
where $\lambda_{\min}$ and $\lambda_{\max}$ are the smallest and the largest eigenvalues of $\Sigma$, respectively. From (\ref{cc}) and $p_3=o(n)$, we have $\|\hat \phi-\phi\|^2=O_p(n^{-1})$.
This implies our claim in (\ref{arg:lim}). 

Note that, 
\begin{eqnarray}
l_n(\phi)=l_n(\hat \phi)+\left.\frac{\partial l_n(\phi)}{\partial \phi^\T}\right|_{\phi=\hat\phi}(\phi-\hat \phi)+\frac{1}{2}(\phi-\hat \phi)^\T\left( \left.\frac{\partial^2 l_n(\phi)}{\partial\phi\partial\phi^\T}\right|_{\phi=\hat \phi}\right) (\phi-\hat \phi)+o_p\left( \frac{1}{n}\right),\nonumber
\end{eqnarray}
which implies that 
\begin{eqnarray}
L_1(\phi\mid \mbox{data})=\psi(\hat \phi\mid \phi, \hat V)(1+o_p(1))=\psi(\hat \phi\mid \phi, V)(1+o_p(1)).\nonumber
\end{eqnarray}

Note that, $$\tilde{p}(z_o|\mbox{data})=\frac{\int  \psi(\hat{\phi}|\phi,V) p(\phi|z_o) p(z_o)   d\phi}{ \int \int \psi(\hat{\phi}|\phi,V)  p(\phi|z) p(z) d\phi dz },$$
and $$ {p}_g(z_o|\mbox{data})=\frac{\int L_1(\phi\mid \mbox{data}) p(\phi|z_o) p(z_o)   d\phi}{ \int \int L_1(\phi\mid \mbox{data})  p(\phi|z) p(z) d\phi dz }.$$ Since we have shown that $L_1(\phi\mid \mbox{data})=\psi ( \hat \phi| \phi, V)\{1+o_p(1)\}$, we thus obtain
\begin{eqnarray*}
	\left|\tilde{p}(z_o|\mbox{data})-p_g(z_o|\mbox{data})\right|\overset{p}{\to} 0,
\end{eqnarray*}
as $n\to \infty$.

\clearpage

\bibliographystyle{abbrvnat}
\bibliography{ref}

\end{document}